27th International Symposium on Superconductivity, ISS 2014

# Single-crystal growth of underdoped Bi-2223

S. Adachi[a], T. Usui[a], K. Takahashi[a], K. Kosugi[a], T. Watanabe[a*], T. Nishizaki[b], T. Adachi[c], S. Kimura[d], K. Sato[d], K. M. Suzuki[d], M. Fujita[d], K. Yamada[e], T. Fujii[f]

[a]*Graduate School of Science and Technology, Hirosaki University, Hirosaki 036-8561, Japan*
[b]*Department of Electrical Engineering and Information Technology, Kyushu Sangyo University, Fukuoka 813-8503, Japan*
[c]*Faculty of Science and Technology, Sophia University, Tokyo 102-8554, Japan*
[d]*Institute for Materials Research, Tohoku University, Sendai 980-8577, Japan*
[e]*Institute of Materials Structure Science, KEK, Tsukuba 305-0801, Japan*
[f]*University of Tokyo, Cryogenic Research Center, Tokyo 113-0032, Japan*

**Abstract**

To investigate the origin of the enhanced $T_c$ ($\approx$ 110 K) of the trilayer cuprate superconductor $Bi_2Sr_2Ca_2Cu_3O_{10+\delta}$ (Bi-2223), its underdoped single crystals are a critical requirement. Here, we demonstrate the first successful in-plane resistivity measurements of heavily underdoped Bi-2223 (zero-resistivity temperatures $\approx$ 20~35 K). Detailed crystal growth methods, the annealing process, as well as X-ray diffraction (XRD) and magnetic susceptibility measurement results are also reported.

*Keywords:* high-$T_c$ superconductor; $Bi_2Sr_2Ca_2Cu_3O_{10+\delta}$ (Bi-2223); single crystal; TSFZ method; annealing; underdoped; resistivity

## 1. Introduction

It is desired that the superconducting transition temperature, $T_c$, of high-$T_c$ cuprates be further increased in order to extend their range of potential applications. It is empirically known that their $T_c$ increases on increasing the number of $CuO_2$ planes in a unit cell, $n$, from $n = 1$ to $n = 3$. However, $T_c$ slightly decreases for $n \geqq 4$ [1, 2]. The microscopic mechanism underlying this behavior is still in dispute. Recently, nuclear magnetic resonance (NMR) measurements revealed [3] that antiferromagnetism and superconductivity coexist in a single $CuO_2$ plane for multilayered ($n \geqq 3$) cuprates in the underdoped region, implying that they are intimately related. Thus, it is important to investigate this anomalous antiferromagnetic metal phase by using transport measurements. The trilayer cuprate superconductor $Bi_2Sr_2Ca_2Cu_3O_{10+\delta}$ (Bi-2223: $n = 3$) is a suitable material for such investigations because it is now available in a single-crystal form [4]. However, controlling this material to a heavily underdoped (HUD) state has been considered difficult. Recently, by using magnetic susceptibility measurements, Wei et al. reported [5] that HUD Bi-2223 can be prepared by annealing in vacuum ($P_{O2} \approx 1 \times 10^{-2}$ Pa) at 500 °C for an extended period (>100 h for $T_c$ < 60 K). However, a similar annealing process often resulted in surface degradation in our samples, and thus, resistivity measurements were unsuccessful.

---

* Corresponding author. Tel.: +81-0172-39-3552; fax: +81-0172-39-3552.
  *E-mail address:* twatana@cc.hirosaki-u.ac.jp

To overcome this problem, we grew single crystals under conditions slightly different from those of an earlier report [4]. Subsequently, the crystals were annealed under conditions milder than those in ref. [5]. Thus, we succeeded in preparing HUD Bi-2223 single crystals and measuring their resistivity.

## 2. Experiment

Bi-2223 single crystals were grown using the traveling-solvent floating-zone (TSFZ) method. Nominal compositions of Bi:Sr:Ca:Cu = (2.25~2.0):(2.0~1.9):2:3 were chosen as the feed-rod compositions. These Bi-rich compositions were expected to reduce the valence of Cu, and thereby the hole concentrations, in comparison with those of the standard composition ratio of Bi:Sr:Ca:Cu = 2.1:1.9:2:3. The feed rod was prepared from powders of $Bi_2O_3$, $SrCO_3$, $CaCO_3$, and CuO (all of purity 99.9% or higher) mixed in the desired ratios and then calcined at 770 ˚C for 12 h in air with repeated regrinding. The calcined powders were hydrostatically pressed under 30 MPa and then sintered at 830 ˚C for 24 h in air. The sintered rod was premelted at a rate of 50~70 mm/h to prepare a dense and thin (~3.0 mm in diameter) feed rod under a mixed gas flow of $O_2$ and Ar at 5 cc/min (10%) and 45 cc/min (90%), respectively. Crystal growth was performed under the same atmosphere as that used for the premelting process. This slightly oxygen-reduced atmosphere was effective in reducing the growth temperature and thus likely elongated the liquidus line. Other growth conditions such as the growth rate (0.05 mm/h) and shaft-rotation speeds (upper: 11 rpm, lower: 10 rpm) were identical to those used in a previous work [4].

The obtained crystals were annealed in two steps. Initial annealing was performed under $P_{O2} \approx 2$ Pa at 600 ˚C for 1~3 h to produce a slightly underdoped crystal ($T_c \approx 85K$) [6]. Next, secondary annealing was performed under $P_{O2} \approx 0.5$ Pa at 500 ˚C for 1~20 h to produce HUD crystals. Note that these secondary-annealing conditions are milder than those in an earlier report [5]. This soft-vacuum condition enabled us to prepare HUD samples without damaging the crystal surface. Furthermore, with this two-step annealing method, we can obtain HUD samples more quickly than with the annealing conditions in ref. [5]. The annealed crystals were characterized using X-ray diffraction (XRD), in-plane resistivity $\rho_{ab}$, and magnetic susceptibility measurements. The $\rho_{ab}$ measurements were performed using the four-terminal DC method.

## 3. Results and Discussion

Figure 1(a) shows a photograph of the entire grown boule, which is homogeneous in diameter because of the constant growth condition (Fig. 1(b)). Figure 1(c) shows a photograph of crystals cleaved from the boule. The grown crystals had large, clean surfaces, as shown in the figure. The typical single-crystal size was $2 \times 2 \times 0.02$ mm$^3$.

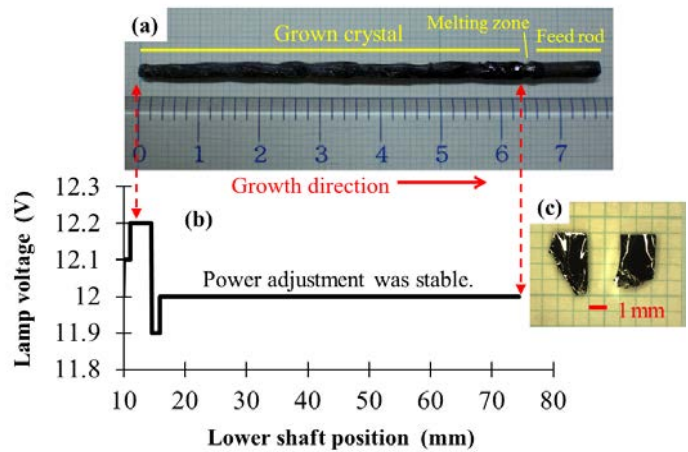

Fig. 1 (color online) (a) Photograph of the grown Bi-2223 ($Bi_{2.25}Sr_2Ca_2C_3O_{10+\delta}$) boule 65 mm in length and approximately 3.0 mm in diameter. (b) Operation history: halogen lamp power vs. lower shaft (seed rod) position. (c) Grown crystals cleaved from the boule.

Figure 2 shows the XRD spectra of the (a) as-grown crystal, (b) crystal annealed in soft-vacuum, and (c) crystal annealed in oxygen. All observed peaks are assigned to the (00l) peaks of Bi-2223. These results indicate that the crystal is a phase-pure Bi-2223 and that no phase separation occurs after the annealing process. The c-axis length is estimated—through a fitting method using the Nelson–Riley (NR) function—as 37.08 Å, 37.14 Å, and 37.01 Å for the data of (a), (b), and (c), respectively. As the c-axis length is known to increase with decreasing oxygen content $\delta$ in Bi-2212 and Bi-2223 [6], this result indicates that excess oxygen is, in fact, removed by annealing under the soft-vacuum condition. Figures 2(d), (e), and (f) show the enlarged (0010) peak of the XRD data and their full widths at half maximum, FWHMs, for (a), (b), and (c), respectively. All the estimated FWHMs have similar values, indicating that the oxygen is rather homogeneously distributed even after the annealing under soft-vacuum conditions. Note that the same sample was used with different treatments for these measurements.

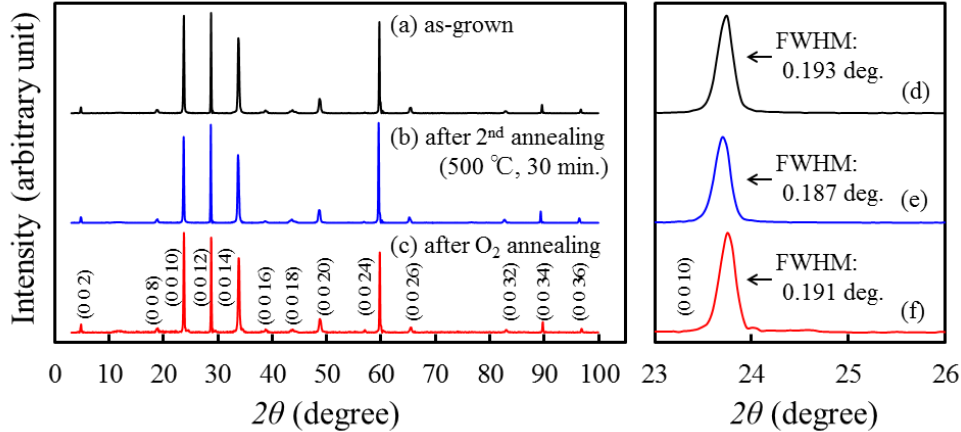

Fig. 2 (color online) Typical XRD patterns using Cu K$_\alpha$ radiation for (a) as-grown, (b) annealed (1$^{st}$: 2 h; 2$^{nd}$:30 min.), and (c) O$_2$-annealed (500 °C 40 h) Bi-2223 (Bi$_{2.25}$Sr$_2$Ca$_2$C$_3$O$_{10+\delta}$) single crystals. All the data were taken using the same sample. (d)-(f) FWHMs of the (0010) peak for (a), (b), and (c), respectively.

Figures 3(a)–(c) show the temperature dependence of the in-plane resistivity, $\rho_{ab}$, for samples A, B, and C, respectively, annealed under soft-vacuum conditions. The secondary-annealing duration was prolonged from 30 min to 50 min for samples A through C. Therefore, the samples are assumed to be underdoped to an increasing degree from sample A to sample C. In fact, the onset temperature, at which the resistivity becomes zero, decreases from 35 K for sample A, to 31 K for sample B, and to 20 K for sample C. In addition, the temperature dependence changes from metallic (samples A and B) to semiconductive (sample C), whereas the magnitude of the resistivity at room temperature increases from sample A to C. All these facts indicate that we succeeded in preparing HUD Bi-2223 single crystals, the doping levels of which are lower than those of previously reported crystals [6].

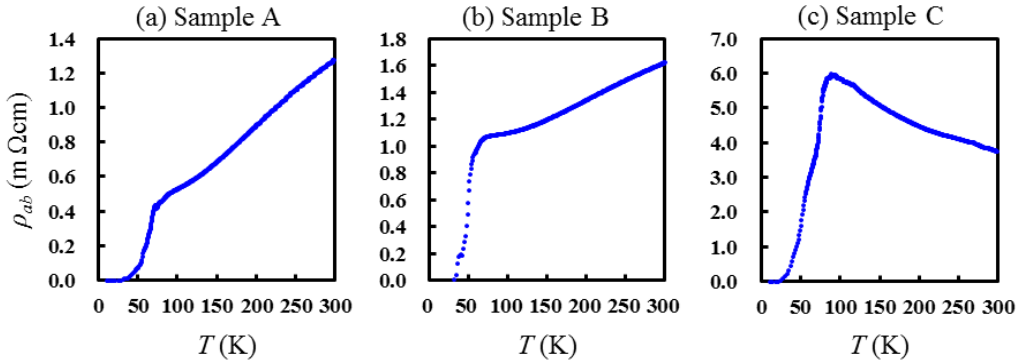

Fig. 3 (color online) (a)–(c) In-plane resistivity, $\rho_{ab}$(T), for HUD Bi-2223 (Bi$_{2.25}$Sr$_2$Ca$_2$C$_3$O$_{10+\delta}$) single crystals. The secondary-annealing durations were (a) 30 min, (b) 40 min, and (c) 50 min. All sample sizes were less than $1 \times 1.5 \times 0.015$ mm$^3$.

Here, we note that the resistive transition curves are considerably broadened. For example, the onset transition temperatures are 71, 55, and 79 K for samples A, B, and C, respectively. This may not be attributed to the oxygen inhomogeneity of the samples (Fig. 2) nor their poor crystallinity, which will be explained below using the magnetic susceptibility data. Rather, we conclude that this is an intrinsic property of underdoped Bi-2223. Recently, an STM/STS study on underdoped Bi-2223 revealed that a new non-dispersive density-of-states (DOS) modulation structure appears with a period 2$a_0$ [7]. The authors of ref. [7] attributed this feature to the multilayer effect of Bi-2223. That is, Bi-2223 has crystallographically inequivalent CuO$_2$ planes: two outer CuO$_2$ planes (OP) with pyramidal oxygen coordination and one inner CuO$_2$ plane (IP) with square oxygen coordination. Since IPs tend to be more underdoped than OPs, a spin stripe, or spin-density wave (SDW), has been thought to be formed on heavily underdoped (probably non-superconducting) IPs. In this case, we can expect electrical decoupling in OPs, which leads to extreme two-dimensional (2D) superconductivity. Therefore, we suppose that the broad resistive transition is caused by the strong superconductive fluctuation originating from this 2D superconductivity. However, to confirm this hypothesis, more detailed testing, such as magnetotransport measurements, will be needed.

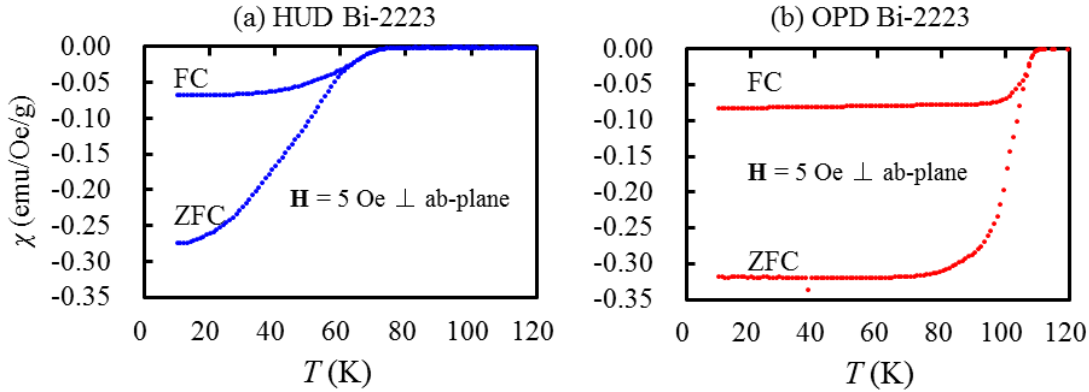

Fig. 4 (color online) Magnetic susceptibility, $\chi$(T), for (a) heavily underdoped (HUD) and (b) optimally doped (OPD) Bi-2223 ($Bi_{2.2}Sr_{1.9}Ca_2C_3O_{10+\delta}$) single crystals. The history of annealing: as-grown crystal → 1st: 3h; 2nd: 12h. (HUD) → $O_2$ 100% 500 ˚C 40 h (OPD).

Figure 4 (a) illustrates the temperature dependence of the magnetic susceptibility $\chi$ of HUD Bi-2223. The data show a very broad superconducting transition ($\Delta T_c \approx 30$ K) with an onset temperature of 70 K. Here, $\Delta T_c$ was determined by the interval between the onset temperature of the Meissner signal and the temperature at which the Meissner signal reached 90% of its maximum value. This result is consistent with the observation of the broad resistive transition (Fig. 3). The HUD Bi-2223 sample was subsequently annealed to the optimally doped (OPD) state, and Fig. 4(b) shows the temperature dependence of the $\chi$ of the resulting sample. In contrast to the HUD case (Fig. 4(a)), the OPD data show a sharp superconducting transition ($\Delta T_c \approx 10$ K) with an onset temperature of 109 K, indicating that the sample quality is good. Therefore, the broad transition observed in the HUD sample cannot be due to the poor crystal quality. Furthermore, the magnitude of $\chi$ at the lowest measured temperature is similar for the HUD and OPD samples, which confirms that the superconductivity is bulk in nature, even in the HUD sample.

## 4. Conclusions

By adopting a Bi-rich feed-rod composition and a slightly oxygen-reduced atmosphere, we succeeded in obtaining large and high-quality Bi-2223 single crystals. Moreover, the obtained crystals were annealed under soft-vacuum conditions, which enabled us to measure the in-plane resistivity because the sample surfaces were free from degradation. The annealed crystals showed very low zero-resistivity temperatures ($\approx$ 20~35 K), indicating that they were in the HUD state. The crystals also showed relatively high superconducting onset temperatures, suggesting that the superconductive fluctuation is enhanced in HUD Bi-2223. This HUD Bi-2223 preparation method can be employed in future studies to gain insight about enhanced $T_c$ in the multilayered (n $\geqq$ 3) cuprates.

## Acknowledgements

We are grateful T. Ito and H. Nakazawa for many helpful discussions. This work was supported by JSPS KAKENHI Grant Number 25400349.